\documentclass[preprint]{revtex4-1}
\usepackage{graphicx}
\usepackage{amsmath}
\usepackage{amssymb}
\usepackage{xspace}
\newcommand{\ket}[1]{\ensuremath{|{#1}\rangle}}

\newcommand{\probe}{\ket{p}}
\newcommand{\target}{\ket{t}}
\newcommand{\pro}{\ket{p}\xspace}
\newcommand{\tar}{\ket{t}\xspace}
\newcommand{\emphsection}[1]{\noindent{\bf #1}\\}

\begin{document}

\title{Probing an Ultracold-Atom Crystal with Matter Waves}
\author{Bryce Gadway}
\author{Daniel Pertot}
\author{Jeremy Reeves}
\author{Dominik Schneble}
\affiliation{Department of Physics and Astronomy, Stony Brook University, Stony Brook, NY 11794-3800, USA} \date{\today}


\begin{abstract}
\textbf{
Atomic quantum gases in optical lattices serve as a versatile testbed for important concepts of modern condensed-matter physics.  The availability of methods to characterize strongly correlated phases is crucial for the study of these systems.  Diffraction techniques to reveal long-range spatial structure, which may complement \emph{in situ} detection methods, have been largely unexplored. Here we experimentally demonstrate that Bragg diffraction of neutral atoms can be used for this purpose. Using a one-dimensional Bose gas as a source of matter waves, we are able to infer the spatial ordering and on-site localization of atoms confined to an optical lattice. We also study the suppression of inelastic scattering between incident matter waves and the lattice-trapped atoms, occurring for increased lattice depth. Furthermore, we use atomic de Broglie waves to detect forced antiferromagnetic ordering in an atomic spin mixture, demonstrating the suitability of our method for the non-destructive detection of spin-ordered phases in strongly correlated atomic gases.
}
\end{abstract}

\maketitle

The diffraction of electron and neutron matter waves from crystalline structures is a standard  tool in crystallography, complementing x-ray techniques~\cite{AshcroftMermin}.  The advent of quantum gases in optical lattices has introduced a new class of highly controllable systems that mimic the physics of solids at lattice constants that are three orders of magnitude larger~\cite{BlochDalibardZwergerReview-2008},  and it appears natural to ask about a possible role for atomic matter-wave diffraction in the characterization of these systems~\cite{KuklovSvistunov-AtomScatt-1999,Sanders-MattWaveScatt-2010}.

Several approaches for detecting the spatial structure of strongly correlated phases of ultracold atoms in optical lattices have so far been taken.  These include the analysis of noise correlations in momentum space~\cite{Folling-SpatialQuantumNoise-2005,Altman-NoiseCorrelations-2004}, accessible after release of the atoms from the lattice, as well as dedicated \emph{in situ} detection methods based on optical imaging~\cite{Gemelke-InSitu-2009,Bakr-SingleAtom-2010,Sherson-SingleAtom-2010} and electron microscopy~\cite{Gericke-EBeamMicroBEC-2008}.  In this context, diffraction experiments have the potential to reveal important information on the existence of long-range order, such as spin-ordered phases in atomic mixtures~\cite{Juha-Janne-OpticalDetection-2003,Cirac-SpinCorrelations-LightScattering-2008,Hulet-DetectingAFM-2010}, in a non-destructive manner and with substantially lower experimental requirements.  Here, optical and atomic matter-wave diffraction are equivalent in the sense that scattering of photons and atoms can be sensitive to both the external and internal state of atomic scatterers. However, there are certain advantages to using matter-wave probes. The de Broglie wavelength of an atomic probe can be tuned freely by controlling its velocity, thus precluding limits on spatial resolution and also providing access to Bragg resonances without the need of varying the angle of incidence. As in the optical case~\cite{Weitenberg-LightScatter-2011,PhotonCounting-2011}, matter-wave probes of high spectral brightness are readily available by using atoms from a Bose--Einstein condensate.

In this work we study the scattering of a \emph{probe} species, consisting of one-dimensional Bose gases, from \emph{target} atoms confined to an optical lattice. For weak confinement, we observe free-particle-like, one-dimensional (1D) collisions~\cite{Olshanii-1Dscattering-1998,Kinoshita-NewtCradle-2006} between the two species, corresponding to inelastic band-structure excitations of the target by the incident probe field. The inelastic scattering is suppressed for more deeply confined target atoms, giving way to elastic scattering and Bragg diffraction, from which the underlying crystalline order can be inferred. We use this matter-wave scattering technique to characterize the introduction of forced antiferromagnetic ordering~\cite{Sengstock-HexLattMixture-2011} in the system.

\vspace{0.1in}

\emphsection{Matter-wave probing of an atomic crystal}
Our experiments begin with a virtually pure Bose--Einstein condensate of $^{87}$Rb atoms in the $|F,m_F\rangle = |1,-1\rangle$ hyperfine ground state, prepared in an optical dipole trap (ODT) of nearly isotropic harmonic trapping frequency $2\pi \times 50$~Hz in the transverse ($x$ and $y$) and vertical ($z$) directions. Along the transverse axes, far-detuned attractive optical lattices ($\lambda_\perp=1064$~nm) are smoothly ramped up in $200$~ms, resulting in an array of one-dimensional Bose gases, with trapping frequency $\omega_z/2\pi \approx 70$~Hz along $z$. The final depths of these transverse lattices, $s_\perp = 40$ (measured in units of the transverse recoil energy $(h/\lambda_{\perp})^2/ 2 m$, where $m$ is the atomic mass), are sufficiently deep so as to energetically restrict all dynamics to the lowest mode of the transverse potential, and to suppress tunneling between the 1D tubes on our experimental timescales.

We next create a mixture of probe (\pro$\equiv|2,-2\rangle$) and target (\tar$\equiv|1,0\rangle$ or $|1,-1\rangle$, see below) species through microwave manipulation of the atoms' internal hyperfine state. By adiabatically ramping up a magnetic field gradient, we then fully separate the probe and target atoms along the longitudinal ($z$) axis. We thereafter smoothly ramp up a state-selective lattice along $z$, formed by light of wavelength $\lambda_z=785$~nm, between the D$_1$ and D$_2$ lines of $^{87}$Rb. This allows (\cite{Deutsch-1998,Pertot-10-FWM,Gadway-BosonicLatticeMixture-2010}, see Methods) for a cancellation of light-shifts (and zero lattice depth) for the probe atoms, while the target atoms experience an attractive lattice along $z$, with period $d=\lambda_z/2$ and variable depth $s_z$ (measured in units of the longitudinal recoil energy $E_{R} = (h/\lambda_{z})^2/ 2 m$). This allows for the target to be driven to a 1D Mott insulator state~\cite{Stoferle-1DMI}, while the probe species remains a superfluid 1D Bose gas. Finally, as detailed below, we examine scattering processes between probe and target at a well-defined relative velocity, $v_{rel}$, relating to a probe de Broglie wavelength of $\Lambda_{dB} = h/m v_{rel}$.

\vspace{0.1in}

\emphsection{1D collisions and inelastic scattering}
We begin by studying collisions between the two species, as the target atoms become localized to a state-selective lattice of increasing depth. In the limit of zero lattice confinement along $z$, the collisions occurring between the \pro and \tar atoms are essentially free-particle collisions, with all scattering restricted to the 1D tubes. Such binary collisions have previously been studied with a single species as a quantum analog of a ``Newton's cradle''~\cite{Kinoshita-NewtCradle-2006}, and with a two-species mixture in the context of damped spin impurity transport~\cite{Palzer-SpinTransport}. Here, we realize such collisions by accelerating the probe atoms into an initially separated sample of target atoms, which is itself at rest.

\begin{figure}[h!]
\centering
\includegraphics[width=3.25in]{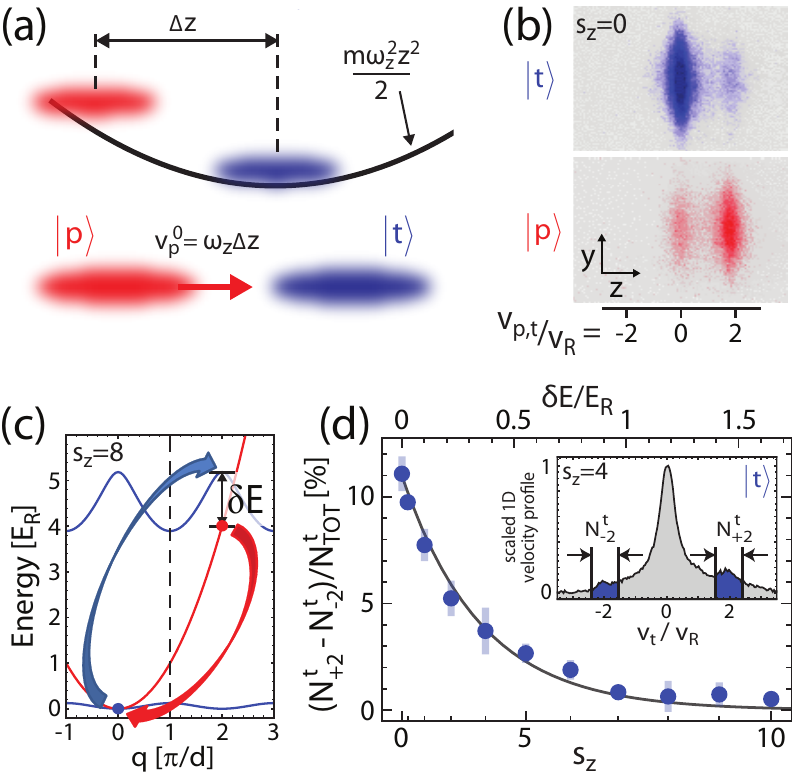}
\caption{Interspecies collisions of one-dimensional bosons.
(\textbf{a}) Probe atoms of species \pro are displaced by a distance $\Delta z$ and then accelerated within a trap potential of trap frequency $\omega_z$ to a final velocity $v_p^\textrm{0} = \omega_z \Delta z$. The \pro atoms then interact with the stationary target atoms \tar at the trap center.
(\textbf{b}) Time-of-flight (TOF) absorption images of the target (top) and probe (bottom) species, with velocity components at $v_{p,t} = 0$ and $2v_R$, due to momentum-exchanging ``Newton's cradle'' (NC) collisions.
(\textbf{c}) Illustration of momentum and energy exchange in the band structure of the optical lattice.
(\textbf{d}) Percentage of target atoms that participate in NC-type collisions. The solid curve is an exponential fit with a decay constant of 0.35~$E_R$. The percentage is determined from the peak asymmetry in the TOF velocity distribution (after summation over the $y$ direction) of the target species, as shown for $v_p^\textrm{0} = 2 v_R$ in the inset, with atom numbers $N^t_{\pm2}$ (peaks) and  $N^t_{\textrm{TOT}}$ (total).
\label{FIG:NewtonsCradle}
}
\end{figure}

As illustrated in Fig.~\ref{FIG:NewtonsCradle}~(a), we set the relative velocity of the collisions by controlling the magnetic field gradient along $z$ that initially separates the two species. The probe atoms experience a longitudinal trapping potential shifted by a distance $\Delta z$, while the target atoms ($|F,m_F\rangle=|1,0\rangle$, insensitive to the applied gradient) remain at the trap center. The magnetic field gradient is quickly extinguished, and after a quarter oscillation in the trap the probe atoms have accelerated to a nominal velocity of $v_p^\textrm{0} \equiv \omega_z \Delta z$ upon interacting with the stationary ($v_t^\textrm{0} =0$) target atoms. At this point, we access their momentum distributions by turning off all confining potentials and separating them by applying a pulsed magnetic field gradient during time of flight (TOF), followed by absorption imaging.

Typical TOF absorption images are displayed in Fig.~\ref{FIG:NewtonsCradle}~(b), for the case of free (along $z$) target atoms and probe atoms incident at $v_{rel} = v_p^\textrm{0} - v_t^\textrm{0} = 2v_R$ ($z$-lattice recoil velocity $v_R = h/2 m d \simeq 5.8$~mm/s). Here, and also for weak $z$-lattice depths, two distinct velocity components at $0$ and $v_{rel}$ can be seen for both species. As the target atoms are initially at rest and the accelerated probe atoms are initially at $v_{rel}$ before colliding with each other, the scattering spectra show energy- and momentum exchanging binary collisions between probe and target atoms, i.e. reflections in the center-of-mass frame of atomic pairs. Whereas reflection and transmission events are indistinguishable for colliding atoms of the same spin~\cite{Kinoshita-NewtCradle-2006}, a spin-mixture gives experimental access to the reflection probability~\cite{Olshanii-1Dscattering-1998}. The atoms collide with a high kinetic energy that far exceeds the mean-field energy of either species, such that the collisions have free-particle character. Using a slightly uneven mixture of the two species (3:2 target to probe atoms), we find that $11\pm1$\% of the target atoms and $14\pm3$\% of the probe atoms are reflected, in fair agreement with the calculated reflection probability of $R=10\%$~\cite{Olshanii-1Dscattering-1998} for our system parameters.

As mentioned above, much effort has been devoted to studying collisions of lattice-free 1D Bose gases, both for the spin-polarized~\cite{Kinoshita-NewtCradle-2006} and spin-mixed cases~\cite{Palzer-SpinTransport}. In our system, we now investigate what happens when the dispersion relations are qualitatively different, with one of the species being subject to a longitudinal optical lattice. To discuss in simple terms our expectations for the case that the target atoms reside in a lattice of finite depth, we describe the momentum and energy exchange in the periodic zone scheme associated with the lattice. For zero lattice depth, the dispersion relations of target and probe atoms coincide, such that the momentum exchange is resonant. However, as illustrated in Fig.~\ref{FIG:NewtonsCradle}~(c), for non-zero lattice depth $s_z\neq0$, each collision demotes an incident probe atom from the first band to the ground band, and vice versa for the target atom involved in the collision. With an energy mismatch $\delta E$ that increases with $s_z$, the momentum-exchange becomes off-resonant, amounting to an inelastic band-structure excitation of the target. This process can occur as long as the energetic uncertainty $\Delta E\sim h/\tau$ associated with the finite interaction time $\tau$ is larger than $\delta E$.

The observed changes of the target's TOF spectra with increasing lattice depth are analyzed in Fig.~\ref{FIG:NewtonsCradle}~(d). In the absence of collisions, $s_z$-dependent, symmetric peaks due to optical diffraction are observed at $v_t=\pm 2 v_R$ as expected (for up to $s_z \sim 20$). Collisions at the incident velocity $v_p^\textrm{0} = 2 v_R$ give rise to an asymmetry between the two peaks, depending on the depth of the lattice. The observed asymmetry decays with the mismatch $\delta E$, in qualitative agreement with our expectation. We estimate the time for probe atoms to traverse half of the target (at which point we switch off the lattice) to be $\tau\sim 10\mu\mbox{m}/2 v_R\simeq 0.9$ms, giving an associated energy uncertainty $\Delta E \sim 0.3 E_R$. This value is in good qualitative agreement with the observed $1/e$ exponential decay constant of 0.35$E_{R}$. Our observation of the suppression of elastic two-body collisions due to band-structure mismatch should be of direct consequence for recently proposed thermometry schemes in strongly correlated systems that are based upon the use of a lattice-free spectator species in contact with lattice-modulated atoms~\cite{McKay-ThermometrySDLatt-2010}.

\vspace{0.1in}

\emphsection{Elastic Bragg diffraction of matter waves}
While inelastic scattering events are seen to die off with increasing lattice confinement, we instead can expect to observe elastic scattering of probe matter waves from the crystal of target atoms. The distribution of elastically scattered probe atoms is determined by the static structure factor $S(q)$ of the target (where $\hbar q$ is the probe momentum transfer). As $S(q)$ is given by the square of the Fourier transform of the target's density, probe scattering can thus reveal information about the underlying spatial distribution of the target atoms. As a caveat, we briefly mention that the study of the dynamic structure factor through low-energy inelastic scattering, which could provide insight into correlations of the target atom array~\cite{Sanders-MattWaveScatt-2010}, is largely precluded in our system due to dominant reflection at low energies (described below), as well as our use of a superfluid probe gas, which itself supports low-energy collective excitations. Such a study may thus be more well-suited to energy-deposition measurements based on optical Bragg spectroscopy~\cite{Clement-BraggScatt-2009,Ernst-BraggSpec-2010}.

The static structure factor, for a deeply modulated target, can be approximated as $S(q)= |\sum_j f_j (q) e^{i q z_j}|^2$ , where the $f_j(q)$ are the amplitudes of isolated scatterers at positions $z_j = j d$, with $j$ an integer. If one assumes them to be identical ($f_j(q)\equiv f(q)$), as in a Mott phase with uniform filling, the structure factor is a product of two terms, $|\sum_j e^{i q z_j}|^2$ and $|f(q)|^2 = \left| \langle \varphi_0 \left| e^{i q z} \right | \varphi_0 \rangle \right |^2$, where $\varphi_0(z)$ describes the on-site density distribution of each scatterer. The first term determines the positions of Bragg diffraction resonances, and reveals the periodicity of the array of scatterers. The second term, the atomic form factor $|f(q)|^2$, or equivalently the single-scatterer envelope function, reveals information of the on-site density distribution.

\begin{figure}[h!]
\centering
\includegraphics[width=3.25in]{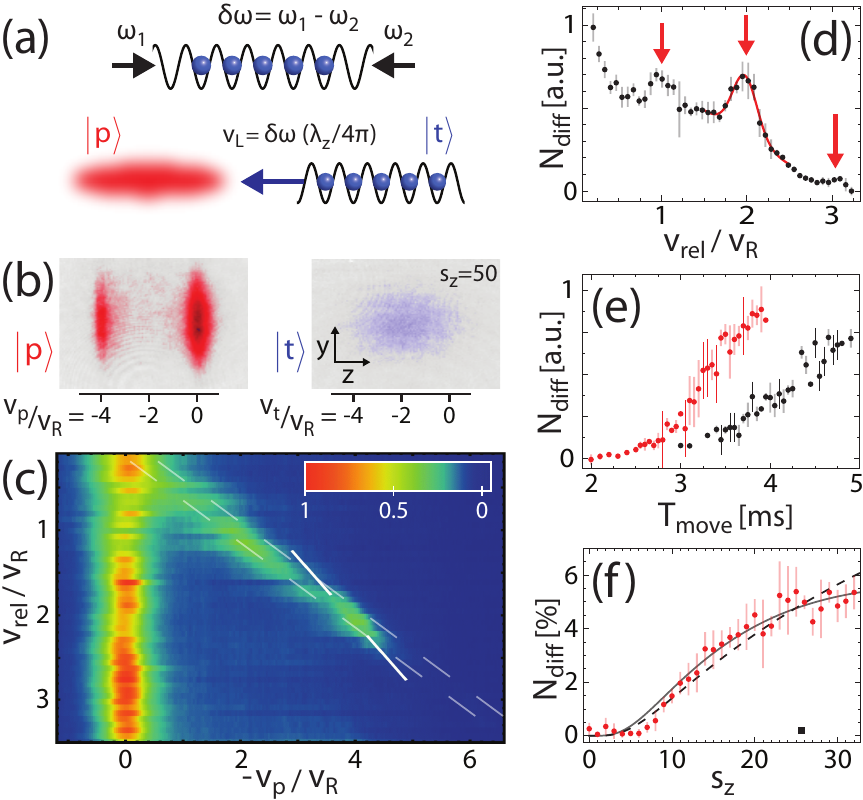}
\caption{Probe scattering from a crystalline target.
(\textbf{a}) Target atoms \tar, strongly confined to a moving state-selective lattice, impinge upon a stationary cloud of probe atoms \pro at a velocity $v_L = \delta \omega (\lambda / 4\pi)$, where $\delta \omega$ is the frequency detuning between the two lattice beams.
(\textbf{b}) TOF images show the \tar and \pro species after interaction.
(\textbf{c}) Probe TOF spectra after interaction with a target at $s_z = 50$, at varied relative velocity $v_{rel}$.  Each horizontal line is obtained from a TOF image as in (b) by integrating along $y$ and normalizing to the total \pro atom number. The solid white lines near $v_{rel}/v_R \sim 2$ illustrate the slope of the clearly visible second-order Bragg resonance. The lines are extrapolations of a linear fit to the positions of maximum out-coupling for $1.8 \lesssim v_{rel} / v_R \lesssim 2.2$.
(\textbf{d}) Dependence of reflection from the crystal on $v_{rel}$. We count the number of atoms scattered to $v_p \sim -2v_{rel}$, i.e. within dashed lines shown in (c). Bragg resonances of the \tar crystal are observed for $v_{rel}/v_R = 1,2,3$, on top of a background contribution due to specular reflection, which decays with increasing $v_{rel}$.
(\textbf{e}) Reflected probe population as a function of the time during which the crystal is moved ($T_{move}$), for fixed velocities $v_{rel}/v_R = 1.5$ (black data) and $v_{rel}/v_R = 2$ (red data). The data fall off at long times due to evolution in the trapping potential.
(\textbf{f}) Percentage of Bragg-diffracted probe atoms (with $v_p^\textrm{0}=2v_R$) as a function of target confinement, $s_z$. The dashed curve is proportional to the form factor $|f(2\times2\pi/d)|^2$ calculated as a function of $s_z$; the solid curve accounts for additional heating effects (see Methods). The black square is a reference measurement without \tar atoms.
Error bars represent one-standard deviation statistical uncertainties, with thin-line error bars in (e) representing extrapolated errors for individual data points.
}
\label{FIG:BraggDiffraction}
\end{figure}

In order to perform crystallographic measurements, we begin by introducing a controlled method to vary the relative velocity (de Broglie wavelength) of matter waves incident on a deeply-confined crystal. Such a wavelength scan is necessary to identify Bragg diffraction resonances, which are expected when a multiple of the probe's de Broglie wavelength coincides with the target's lattice spacing (i.e. $2 d = n \Lambda_{dB}$, or $v_{rel} = n v_R$, with $n$ the order of diffraction), and more generally it allows study of the dependence of scattering on the probe's de Broglie wavelength. As illustrated in Fig.~\ref{FIG:BraggDiffraction}~(a), the target atoms (here in the $|1,-1\rangle$ state) are loaded into a very deep state-selective lattice initially at rest in the lab frame. This lattice is then made to move at a well-defined velocity $v_L = \delta \omega (\lambda_z / 4\pi)$ by detuning the relative frequency of the two interfering laser beams that create the lattice by an amount $\delta \omega$. The pinned target atoms, which are initially far-separated from the probe atoms, follow the moving lattice minima at a variable velocity $v_{rel}$ for a total time $T_{move}$, including a set delay time before the probe and target make contact. The large initial separation helps to avoid the formation of a quantum emulsion~\cite{Roscilde-QuantEmuls-2007}, resulting when homogeneous mixtures are loaded into state-dependent lattice potentials, which can affect interspecies transport properties~\cite{Gadway-BosonicLatticeMixture-2010,Gadway-BoseGlass-2011}.

In Fig.~\ref{FIG:BraggDiffraction}~(b), we show TOF absorption images of both species following interaction, taken for the case of deep confinement ($s_z = 50$) and a velocity $v_{rel} = 2v_R$. The target's velocity distribution is very broad, as expected for a deep 1D Mott state, and is centered around the velocity of the moving lattice ($-2v_R$). The velocity distribution of the probe, initially centered around $v_p = 0$, displays a peak of atoms transferred to $-4v_R$. A series of such scattering spectra (integrated along the transverse $y$ axis) is shown in Fig.~\ref{FIG:BraggDiffraction}~(c), for a range of relative velocities $v_{rel}$. For each velocity, the time $T_{move}$ is chosen so that the target crystal enters the probe and then moves for $1.4$~ms (kept much less than the trap period $T_z = 2 \pi / \omega_z \sim 14$~ms to avoid the dispersion of out-coupled atoms to different velocities due to propagation in the trapping potential). From the scattered probe spectra as in Fig.~\ref{FIG:BraggDiffraction}~(b) we observe that, in addition to the line of ``transmitted'' atoms at $v_p = 0$, some probe atoms are out-coupled (i.e. reflected) due to the target crystal. The reflection of probe atoms comes from two apparent elastic mechanisms - specular reflection and resonant Bragg diffraction. We note that both of these mechanisms are to be distinguished from observations of free-space four-wave mixing in two-component mixtures~\cite{Pertot-10-FWM,McKay-ThermometrySDLatt-2010}, which does not persist when one species loses matter-wave coherence.

The specular contribution is due to the reflection of probe atoms from the potential ``step'' of the target crystal (energy mismatch outside and inside the crystal). This reflection is dominant at low velocities and gives way to transmission at larger probe velocities (kinetic energies), where the band structure of the target ``atomic crystal'' is nearly free-particle-like. The specular contribution leads to the transfer of probe atoms to $v_p = -2 v_{rel}$ and shows no resonant structure. In contrast, Bragg reflection from scattering centers of the target crystal occurs at values $v_{rel}/v_R = n$ and results in the transfer of probe atoms to velocities $v_p = -v_{rel} - n v_R$. Such a ``locking'' to a Bragg resonance is observable through a change in the slope of the out-coupled branch or probe atoms. Results of the velocity scan, which displays these features, are shown in Fig.~\ref{FIG:BraggDiffraction}~(c).

In Fig.~\ref{FIG:BraggDiffraction}~(d) we count the number of probe atoms out-coupled to $v_p \approx 2 v_{rel}$, and plot it as a function of $v_{rel}/v_R$. Three resonant peaks in the number of reflected atoms are observed for $v_{rel}/v_R = 1,2,3$, corresponding to first-, second-, and third-order Bragg processes. In addition to these resonances, a significant non-resonant contribution due to specular reflection is observed, which as expected decays with increasing velocity $v_{rel}$. We fit the second-order resonance peak with a Gaussian on top of a linear slope [red line in Fig.~\ref{FIG:BraggDiffraction}~(d)], and extract a $1/\sqrt{e}$-width of $2\sigma_v = 0.3 v_R$, reflecting the in-trap velocity width of the probe species. The width is a factor of two less than the directly observed TOF velocity width of the probe atoms~\cite{Fabbri-Bragg-Fluctuating1D}, likely owing to effects of interaction-induced expansion~\cite{Stenger-Bragg-1999}.

Using our ability to vary the total time $T_{move}$ that the target crystal moves at constant velocity, we can also study the temporal buildup of the out-coupled probe population. In contrast to optical scattering, this may in the future be used to provide ``line-of-sight'' information in matter-wave scattering experiments. In Fig.~\ref{FIG:BraggDiffraction}~(e), we plot the temporal buildup of probe population to $v_p \sim -2v_{rel}$, for the cases $v_{rel} = 1.5 v_R$ and $2v_R$. The difference between incoherent (specular reflection) and coherent (Bragg) processes should lead to differences in growth behavior. Indeed, there are some indications of a more nonlinear initial growth for $v_{rel} = 2v_R$ than for $v_{rel} = 1.5v_R$ case. However, since the relatively short coherence length of the probe [$L \sim \hbar / (m \sigma_v) =0.8$~$\mu$m] precludes a fully coherent temporal evolution, the observed growth will be mostly determined by the time-dependent overlap of the inhomogeneous profiles of the probe and target.

Finally, we recall the dependence of the static structure factor $S(q)$ on the atomic form factor $|f(q)|^2$, in relation to the three Bragg resonances observed in Fig.~\ref{FIG:BraggDiffraction}~(d). In the harmonic approximation, the form factor is proportional to $\exp(-q^2\sigma_z^2/2)$, where $\sigma_z$ is the harmonic-oscillator length characterising the extent of the atomic distribution on each site. While the third-order peak ($q=3$) is smaller than the second-order peak ($q=2$), the expected momentum dependence is partially masked, owing to a fixed probe-target interaction time that results in the probe atoms of lower velocity (such as for the first-order peak) interacting with fewer scattering centers of the target crystal. However, we can directly probe the contribution from the atomic form factor, which can be seen to be formally identical to the Debye--Waller factor ($\exp(-q^2 \langle u^2 \rangle / 2)$) describing the reduction of elastic scattering due to position fluctuations $\langle u^2\rangle$ of scatterers in an ionic crystal~\cite{AshcroftMermin}. The larger $\sigma_z$, the smaller $|f(q)|^2$, in analogy to the decrease of the Debye--Waller factor with temperature. We point out, however, that in our case these fluctuations are not thermal but arise from zero-point motion, and may be tuned through $\sigma_z$ by varying $s_z$.

Using the earlier described method of probe acceleration in the longitudinal trapping potential (with $|1,0\rangle$ target atoms as before), we can study the dependence of the second-order Bragg diffraction amplitude on lattice depth $s_z$, as shown in Fig.~\ref{FIG:BraggDiffraction}~(f). The plot shows an increase of diffraction intensity as the target atoms become more localized, and the data is seen to be in good qualitative agreement with the calculated dependence of $|f(q)|^2$ on $s_z$ (using Bloch functions for $\varphi_0$). Even better agreement can be obtained by taking into account $s_z$-dependent heating due to Rayleigh scattering from the laser beams forming the longitudinal lattice (see Methods).

\vspace{0.1in}

\begin{figure}[h!]
\centering
\includegraphics[width=3.25in]{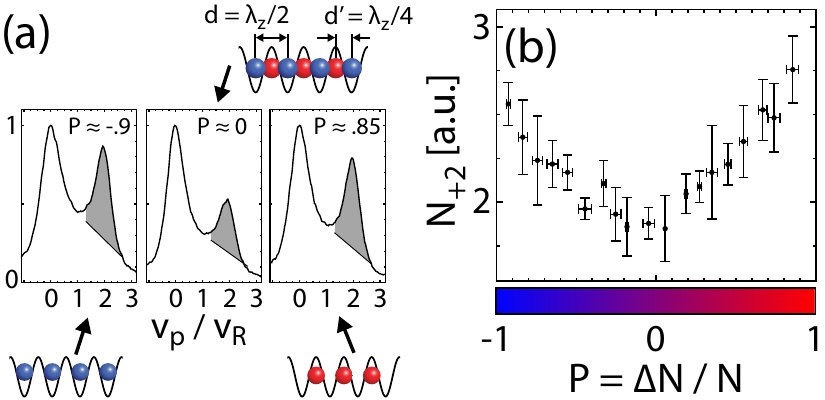}
\caption{Detecting forced antiferromagnetic order via matter-wave scattering.
(\textbf{a}) Velocity distributions of scattered probe atoms following interaction with a moving (at $v_{rel}=-v_R$) crystal of atoms in a tunable spin-mixture of red $|r\rangle$ and blue $|b\rangle$ states, attracted to and repelled from intensity maxima of a state-dependent lattice, respectively. Shown are the cases of a crystal of nearly all $|b\rangle$ atoms ($P \sim -1$), nearly all $|r\rangle$ atoms ($P \sim 1$), and a balanced mixture of the two states ($P \sim 0$).
(\textbf{b}) The number of Bragg-diffracted probe atoms [shaded regions in (a)], as a function of the spin population imbalance of the target crystal, $P = \Delta N / N$. A distinct minimum is observed for $P \approx 0$, relating to the case of a balanced spin-mixture crystal with maximal forced antiferromagnetic order at spacing $d' = \lambda_z / 4$.
}
\label{FIG:ForcedAFM}
\end{figure}

\emphsection{Detecting forced antiferromagnetic ordering}
To further demonstrate that matter-wave diffraction may be used to probe non-trivial structure, we show here that it gives access to a clear signature of forced antiferromagnetic ordering~\cite{Sengstock-HexLattMixture-2011} in a mixed-spin crystal. As before, we start with a mixture of atoms in one-dimensional tubes, with a probe that sees no longitudinal lattice along $z$ (\pro, here $|1,1\rangle$ atoms, with $\lambda_z \sim 788$~nm). However, instead of using only one additional species that experiences an attractive lattice along $z$, our target now consists of two species. The first one $|r\rangle \equiv |1,-1\rangle$ again experiences an attractive lattice potential, whereas for the second species $|b\rangle \equiv |2,-2\rangle$ the potential is repulsive. Thus, the $|r\rangle$ atoms will be drawn to the intensity maxima, while the $|b\rangle$ atoms will be forced to the minima, as illustrated in Fig.~\ref{FIG:ForcedAFM}~(a). While we keep the total population of these two species fixed, we vary their relative population, which can be quantified by the net ``polarization'' of the crystal $P = \Delta N / N$, with $\Delta N = N_r - N_b$ and $N = N_r + N_b$.

When the target is ``spin-polarized'' consisting of either all $|r\rangle$ or all $|b\rangle$ atoms, the situation is as before, with a lattice constant of $d = \lambda_z / 2 \sim 400$~nm. However, when both species are present, the new dominant length scale, between atoms of the two species sitting on distinct sets of sites, will be given by $d' = \lambda_z / 4 \sim 200$~nm. This may be viewed as a crystal with the original periodicity and a two-atom basis, or as a new crystal structure having half the lattice period, given that the interactions of the probe with the $|r\rangle$ and $|b\rangle$ atoms are approximately the same (see Methods). This change results in a different diffraction spectrum for the matter wave probes. With the addition of a second species, as scattering centers separated by $d$ give way to those of a smaller spacing $d'$, the original first-order diffraction peak at $v_{rel}/v_R = -1$ will become diminished, and entirely disappear if an equal mixture of the two species uniformly fills the lattice, with a new first-order peak now occurring at half the momentum-space frequency, $v_{rel}/v_R = -2$.

To probe the mixed-spin crystal, we move it at a constant velocity of $v_{rel}/v_R = -1$ with respect to the probe atoms, for a fixed interaction time of $1$~ms. In Fig.~\ref{FIG:ForcedAFM}~(a), we show probe velocity distributions for the three cases $P \sim -1$, 0, and 1. As can be seen, the number of probe atoms out-coupled to $2 v_R$ is much lower for the spin mixture ($P=0$) than for either of the nearly spin-polarized cases. We note that an appreciable number of out-coupled atoms appears even for the balanced spin-mixture, which is most likely due to specular reflection as in the previously studied case, while the presence of unoccupied sites of either the attractive or repulsive lattices may also cause some Bragg diffraction consistent with the original lattice spacing.

To characterize how the crystal structure changes as the population imbalance is continuously tuned, we count the number of probe atoms transferred to a velocity region around $v_p \sim 2 v_R$. As a function of the crystal ``polarization'', the transferred population shows a distinct minimum near $P=0$ for a balanced mixture, as shown in Fig.~\ref{FIG:ForcedAFM}~(b). This example readily demonstrates how matter-wave scattering can be used to detect changes to the crystal structure of an ultracold lattice gas, and through species-selective scattering may eventually be used to detect quantum-magnetic spin-ordered states. Finally, we point out that the full tunability of the wavelength for atomic matter waves should provide a distinct advantage over optical scattering in certain circumstances, namely in studying density structures not formed in optical lattice potentials, such as through self-organization or with magnetic trapping potentials. By using probes of small de Broglie wavelength, this could also include the study of novel quantum states realized at more easily attainable temperatures in systems with smaller characteristic length scales. For example, at fixed lattice depth $s$ and tunneling-to-interaction ratio $t/U$ (e.g., by control of the scattering length $a_s$), the N\'{e}el temperature $T_N$, marking the onset of antiferromagnetic ordering, scales with lattice spacing $d$ as $T_N \propto d^{-1/2}$~\cite{CoolingReview-McKay}, and has an even more favorable scaling for fixed $a_s$ and variable $s$.

\vspace{0.1in}

\emphsection{Conclusion.}
In this work, we have demonstrated that matter-wave diffraction can be used to characterize the crystalline structure of strongly correlated atoms in an optical lattice. In the future, these techniques may be extended to the characterization of various novel states of ultracold matter, such as charge- and spin-density waves, magnetically-ordered ground states of quantum gas mixtures, and even self-assembled structures such as Tonks--Girardeau gases of fermionized bosons, Abrikosov vortex lattices~\cite{Madison-Dalibard-VortexNucleation-2001,Raman-VortexNucleation-2001}, and dipolar crystals~\cite{Pupillo-DipolarCrystal}.
%
%

\vspace{0.1in}
{\centering\emphsection{Methods}}
\vspace{0.1in}

\small{\textbf{Mixture preparation and detection.}}
\footnotesize{
Starting with an optically-trapped Bose--Einstein condensate of $^{87}$Rb atoms in the $\ket{F,m_F} = \ket{1,-1}$ hyperfine ground state, the system is split into an array of one-dimensional tubes in the horizontal $xy$ plane. Next, at a magnetic bias field of 1.7~G along $z$, we create hyperfine state mixtures via combinations of microwave Rabi pulses and Landau--Zener sweeps. Different combinations of hyperfine states are used to perform particular experiments on interspecies scattering, with slight variations in the mixture characteristics. For detection, all species are absorptively imaged on the $F=2 \to F'=3$ cycling transition, concurrent with optical pumping from $F=1\to F'=2$.

The initial two-species mixture of probe $\probe \equiv \ket{2,-2}$ and target $\target \equiv \ket{1,0}$ atoms, used to study the $s_z$-dependent scattering [Fig.~\ref{FIG:NewtonsCradle} ; Fig.~\ref{FIG:BraggDiffraction}~(f)], contains a total of $(1.6 \pm 0.2) \times 10^5$ atoms, with $60\%$ of atoms in the target state. For this mixture, a final magnetic bias field of 7.4~G is employed to suppress $\ket{1,0} \leftrightarrow \ket{1,\pm 1}$ spin-changing collisions. A second binary mixture of probe $\probe \equiv \ket{2,-2}$ and target $\target \equiv \ket{1,-1}$ atoms is used in conjunction with the moving optical lattice [Fig.~\ref{FIG:BraggDiffraction}~(a--e)]. This mixture contains a total of $(2.8 \pm 0.5) \times 10^5$ atoms, with $33\%$ target atoms. A three-species mixture of one probe species ($\probe \equiv \ket{1,1}$) and two target species ($\ket{2,-2}$ and $\ket{1,-1}$) is used to study forced antiferromagnetic ordering [Fig.~\ref{FIG:ForcedAFM}]. This mixture contains a total of $(1.5 \pm 0.2) \times 10^5$ atoms, with $50\%$ of the atoms in the target. The target, consisting of two different species, has a fully tunable spin composition [c.f. Fig.~\ref{FIG:ForcedAFM}~(b)]. The intra- and interspecies scattering lengths for all the states used are approximately equal to $5.3$~nm, i.e. the background scattering length for $^{87}$Rb atoms.
}

\small{\textbf{One-dimensional tubes.}}
\footnotesize{
The array of one-dimensional tubes is created by ramping up two orthogonal lattices (along $x$ and along $y$) within 200~ms to depths of $s_\perp = 40$, via partial retroreflection of the ODT's laser beams~\cite{Pertot-09-JPB}, resulting in a final trapping frequency $\omega_z / 2 \pi \approx 70$~Hz along $z$ as determined by dipole oscillations. In the harmonic approximation, the transverse oscillation frequency in each tube is $\omega_\perp /2 \pi = 26$~kHz, resulting in a spacing to the first allowed transverse excited mode of $2 \hbar \omega_{\perp} \approx 14 \ E_{R}$, which is greater than the probe kinetic energy $T = (v_p^\textrm{0} / v_R)^2 E_R$ and all other relevant energy scales (thermal as well as all intra- and interspecies interaction energies). For all the cases studied the probe species, while 1D, is not deeply within the Tonks--Girardeau regime~\cite{Girardeau-BoseFermiCorr-1960}, and is characterized by a Lieb--Liniger parameter value $\gamma \lesssim 1$~\cite{LiebLiniger-1963,Olshanii-TFtoTonks-PRL-2001}
}

\small{\textbf{State-selective lattice potential.}}
\footnotesize{
The state-selective lattices are formed by interfering two laser beams ($1/e^2$ radius $\sim 230~\mu$m , the same polarization) with tunable wavelength between the $^{87}$Rb D$_1$ and D$_2$ lines to effect a light-shift cancellation for the probe atoms. In the case of a stationary target of $|1,0\rangle$ atoms, the lattice is loaded to a variable depth $s_z$ with an s-shaped curve in 75~ms and held for an additional 5~ms prior to acceleration of the probe atoms. This lattice is made from fully retroreflected laser light of wavelength $\lambda=785$~nm of $\sigma^-$ polarization. For the case of a moving target of $|1,-1\rangle$ atoms (wavelength and polarization as in the stationary case), the lattice is first smoothly loaded in 45~ms to a depth of $s_z = 10$, exceeding the critical depth of the 1D Mott insulator transition, and then loaded in 5~ms to a depth of $s_z = 50$ to freeze the atoms to the sites of the lattice. The lattice is then moved, by introducing a relative frequency detuning of $\delta \omega$ between the forward and retroreflected laser beams comprising the $z$-lattice, via two acousto-optic modulators driven by phase-locked function generators.

For the case of a spin-mixed crystal with forced antiferromagnetic ordering, the two species are quickly loaded into the state-dependent lattice in 2~ms following creation of the target mixture to avoid spatial separation in a small magnetic field gradient used to separate the probe. The lattice has modulation depths $s_z = 5$ and $s_z = 12$, for the $|b\rangle = |2,-2\rangle$ (repulsive lattice) and $|r\rangle = |1,-1\rangle$ (attractive lattice) atoms, respectively. This lattice is formed with light of wavelength $\lambda \sim 788$~nm, of slightly elliptical polarization. After loading, this lattice is moved at a fixed velocity as above, with a restriction to relatively low velocities to ensure that both target species faithfully follow the optical potential at these modest depths, as observed in the velocity distributions of the target species.

All the lattice depths are calibrated using Kapitza--Dirac atom diffraction~\cite{Gadway-KD-2009}, with a systematic uncertainty of about $3 \%$. For probe atoms, the lattice potential is sufficiently ``zeroed'' even for greatest available optical potentials such that no diffraction is observed for a pulsed-on lattice or for application of a linear potential gradient while the lattice is present (i.e. no Bloch oscillations). To further ``zero'' the optical lattice, the probe atoms are loaded into 1D tubes, and we either minimize excitations of probe atoms undergoing dipole oscillations, or in the case of a moving lattice we move with velocity of $v_L = v_R$, with no observable transfer of probe atoms to non-zero velocities.
}

\small{\textbf{Second-order Bragg peak population vs. $\mathbf{s_z}$.}}
\footnotesize{
The percentage of second-order Bragg-reflected probe atoms $\textrm{N}_{\textrm{diff}}$ plotted in Fig.~\ref{FIG:BraggDiffraction}~(f) is determined from a fit of the TOF spectrum with Gaussian peaks for (1) the transmitted probe around $v_p \approx v_p^{\textrm{0}}$ ; (2) a broad incoherent background centered at $v_p = 0$ ; (3) Bragg-reflected atoms at $v_p = -2 v_R$. $\textrm{N}_{\textrm{diff}}$ is determined as the amount of Bragg-reflected atoms normalized with respect to the total probe population.
}

\small{\textbf{Heating effects.}}
\footnotesize{
The observed percentage of Bragg diffracted atoms is in qualitative agreement with that expected from the form factor $| f (q) |^2$, up to an overall scaling factor [dashed line in Fig.~\ref{FIG:BraggDiffraction}~(f)]. However, the observed signal saturates at large $s_z$ while the expected curve does not. We attribute this mostly to a heating of the probe species, leading to reduced coherence and an increased spectral width, as observed through a linear increase in the probe's TOF velocity-width with increased hold time of a deep lattice ($s_z = 25.5$) prior to probe acceleration to $v_p^\textrm{0} = 2v_R$. The Bragg diffraction signal exhibits a roughly exponential decay with hold time($1/e$--time $t_{\textrm{heat}} \sim 150-200$~ms). Assuming that contributions to this from heating due to Rayleigh scattering should scale as the time integral over the lattice depth, we expect a correction factor of $\exp(-0.01 s_z)$ to modify the form factor. A fit to the data with $| f (q) |^2 \times \exp(\beta s_z)$ yields $\beta = -0.015$ [value used for the dashed curve in Fig.~\ref{FIG:BraggDiffraction}~(f)], consistent with the role of such heating.
}

\vspace{0.1in}
\emphsection{Acknowledgements.}
\vspace{0.1cm}
We thank G.~Pupillo and K.~Le~Hur for discussions, and M.~G.~Cohen and T.~Bergeman for valuable comments on the manuscript. This work was supported by NSF (PHY-0855643), and the Research Foundation of SUNY. B.G. and J.R. acknowledge support from the GAANN program of the US DoEd.

%
%
%



\clearpage

\clearpage

\clearpage

\end{document}